\documentclass[%
 reprint,
nofootinbib,
nolongbibliography,
 amsmath,amssymb,
 aps,
prl,
]{revtex4-2}
\usepackage[left=1in,right=1in,top=1in,bottom=1in]{geometry}
\usepackage[%
  colorlinks=true,
  urlcolor=blue,
  linkcolor=blue,
  citecolor=blue
]{hyperref}
\usepackage{cancel}
\usepackage{graphicx,sidecap}
\graphicspath{{figs/}}
\usepackage{enumerate}
\usepackage{color}
\usepackage{ulem} 

\usepackage{dcolumn}
\usepackage{bm}
\usepackage{amsmath,amssymb}
\usepackage{aas_macros}
\usepackage{caption}
\usepackage{subfig}

\usepackage{siunitx}
\DeclareSIUnit{\parsec}{pc}
\DeclareSIUnit{\dmunit}{pc cm^{-3}}

\let\oldthebibliography\bibliography
\renewcommand{\bibliography}[1]{%
  \oldthebibliography{#1}
  \let\oldbibitem\bibitem
  \let\oldtextsc\textsc
  \def\oldbbland{et}
  \newcounter{authorcount}
  \def\bibitem[##1]##2{%
    \let\textsc\oldtextsc
    \let\bbland\oldbbland
    \oldbibitem[##1]{##2}%
    \let\textsc\mytextsc%
    \let\bbland\mybbland
    \setcounter{authorcount}{0}
  }
  \def\mybbland{\setcounter{authorcount}{0}\oldbbland}
  \def\dropetal##1.{ \bbletal}
  \def\mytextsc##1{%
    \oldtextsc{##1}%
    \stepcounter{authorcount}%
    \ifnum\value{authorcount}=5\relax%
      \expandafter\dropetal%
    \fi%
  }%
}

\usepackage[mathlines]{lineno}

\newcommand{\msbold}[1]{#1}

\date{\today}

\begin{document}

\preprint{APS/123-QED}

\title{Nulling baryonic feedback in weak lensing surveys \\ using cross-correlations with fast radio bursts} 

\author{Calvin Leung}
\email{calvin\_leung@berkeley.edu}
\affiliation{
    Miller Institute for Basic Research, University of California Berkeley, Berkeley, CA, 94720, USA
}
\affiliation{
    Department of Physics, University of California Berkeley, Berkeley, CA, 94720, USA
}
\affiliation{
    Department of Astronomy, University of California, Berkeley, CA 94720, USA
}

\author{Josh Borrow}
\affiliation{
    Department of Physics and Astronomy, University of Pennsylvania, 209 South 33rd Street, Philadelphia, PA 19104, USA
}
\author{Kiyoshi W. Masui}%
\affiliation{%
 MIT Kavli Institute for Astrophysics and Space Research, 70 Vassar Street, Cambridge, MA 02139
}%
\affiliation{
    Department of Physics, Massachusetts Institute of Technology, 77 Massachusetts Ave, Cambridge, MA 02139, USA
}
\author{Shion Andrew}%
\affiliation{%
 MIT Kavli Institute for Astrophysics and Space Research, 70 Vassar Street, Cambridge, MA 02139
}%
\affiliation{
    Department of Physics, Massachusetts Institute of Technology, 77 Massachusetts Ave, Cambridge, MA 02139, USA
}
\author{Kai-Feng Chen}%
\affiliation{%
 MIT Kavli Institute for Astrophysics and Space Research, 70 Vassar Street, Cambridge, MA 02139
}%
\affiliation{
    Department of Physics, Massachusetts Institute of Technology, 77 Massachusetts Ave, Cambridge, MA 02139, USA
}
\author{Joop Schaye}
\affiliation{Leiden Observatory, Leiden University, PO Box 9513, NL-2300 RA Leiden, The Netherlands}
\author{Matthieu Schaller}
\affiliation{Lorentz Institute for Theoretical Physics, Leiden University, PO Box 9506, NL-2300 RA Leiden, The Netherlands}
\affiliation{Leiden Observatory, Leiden University, PO Box 9513, NL-2300 RA Leiden, The Netherlands}

%

\date{\today}

\begin{abstract}
    Baryonic feedback is a leading contaminant in studying dark matter using cosmic shear. 
    This has meant omitting much of the data during cosmological inference, or forward-modeling the spatial distribution of gas around dark matter halos using models for baryonic feedback, which introduces nuisance parameters and raises questions of potential model misspecification.
    We propose a novel method of ``baryon nulling'' using cross-correlations between shear maps and fast radio burst (FRB) dispersion measures. 
    We use the FLAMINGO suite of hydrodynamical simulations, whose power spectra span a wide yet realistic range of feedback variations, to demonstrate that our nulling procedures reduces sensitivity to feedback at $k \approx 1$ Mpc$^{-1}$ by a factor of a few, offering a valuable model-independent alternative to standard forward modeling procedures, which often assume a family of feedback models.
    Sensitive FRB surveys such as CHORD and the DSA-2000, and wide area surveys like Rubin, Roman, and Euclid, may enable the megaparsec-scale clustering of dark matter to be precisely constrained without a baryonic prior.
\end{abstract}

\maketitle


Weak gravitational lensing is a direct probe of dark matter through the total matter power spectrum. 
Cosmological parameter constraints from Stage III cosmic shear measurements such as the Kilo-Degree Survey~\citep{kuijken2015gravitational,li2023kids,tr_oster2022joint}, Subaru-HSC~\citep{aihara2018hyper,dalal2023hyper,li2023hyper}, and the Dark Energy Survey~\citep{amon2022dark,secco2022dark} have paved the way for LSST~\citep{tyson2002large}, Roman~\citep{eifler2021cosmology}, and Euclid~\citep{scaramella2022euclid}.

Cosmic shear is sensitive to the total matter power spectrum on a mixture of spatial scales ($k \sim 0.01-10$ Mpc$^{-1}$), but astrophysical processes like gas cooling~\citep{white2004baryons,jing2006influence,rudd2008effects}, AGN feedback~\citep{levine2006active,duffy2010impact,mccarthy2010case,van_daalen2011effects}, and supernova feedback~\citep{dalla_vecchia2008simulating,van_daalen2011effects} modify the matter power spectrum at small scales ($k \gtrsim 1$ Mpc). 
These astrophysical processes---collectively called ``baryonic feedback''---already complicate the interpretation of cosmic shear power spectra (see~\citet{chisari2025mapping} for a recent review) and limit the constraining power of KiDS, DES, and HSC data~\citep{tr_oster2022joint,aric_o2023des,terasawa2024exploring,chisari2025mapping}. 
Conversely, mitigating the impact of baryonic feedback can unlock the full constraining power of smaller angular scales in next-generation surveys to e.g. better constrain the sum of the neutrino masses.

Fast radio bursts -- millisecond-duration radio transients originating at cosmological distances -- can play an important role in understanding baryonic feedback. While the emission mechanisms and origins of FRBs remain unknown~\citep{petroff2022fast}, they are abundant enough ($\sim 10^4$ Gpc$^{3}$ yr$^{-1}$; see~\citealp{shin2023inferring}) to be used as statistical tracers of the cosmic gas distribution through their dispersion measures, which probe the electron distribution. 

Modern hydrodynamical simulations such as EAGLE, Illustris, IllustrisTNG, and MillenniumTNG~\citep{schaye2015eagle,nelson2015illustris,nelson2019illustristng,pakmor2023millenniumtng}, BAHAMAS~\citep{mccarthy2017bahamas}, and FLAMINGO~\citep{schaye2023flamingo} show that the gas distribution is more sensitive to feedback than the stellar component of galaxies.
This makes cross-correlation of weak lensing shear maps with FRB dispersion measures a promising way to constrain feedback, as proposed in~\citet{reischke2023calibrating}. FRB x weak lensing cross-correlations are complementary to X-rays ~\citep{ferreira2023xray} and tSZ~\citep{vikram2017measurement,amodeo2021atacama} cross-correlations since those gas probes are most sensitive to hot gas, while FRBs probe all ionized gas. 

In this paper, we propose to carry out weak lensing cosmology using a new observable that is insensitive to the large-scale baryon distribution.
Our ``baryon-nulled'' power spectrum uses cross-correlations between cosmic shear and fast radio burst dispersion measures to nullify the baryonic contributions to the total matter power spectrum.
This suppresses the impact of feedback on the matter power spectrum to first order in $\Omega_b /\Omega_c$, where $\Omega_b \approx 0.05$ and $\Omega_c \approx 0.25$ are the average baryon (b) and cold dark matter (c) densities as a fraction of $\rho_\mathrm{crit} = 3 H_0^2 / (8\pi G)$.
The idea is as follows.
\msbold{We express density fluctuations for a tracer $i$ as $\delta_i = (\rho_i - \overline{\rho}_i)/\overline{\rho}_i$, where $\overline{\rho}_i$ is the mean density of the tracer, and define the mass fraction of the tracer as $f_i = \Omega_i / \Omega_m$.}
\footnote{We omit massive neutrinos from the total and dark matter fields since $\Omega_\nu/\Omega_c \ll 1$. Note that all hydrodynamical simulations and their dark-matter-only counterparts in this paper assume the D3A cosmology with $\sum m_\nu = 0.06$ eV~\citep{schaye2023flamingo}; the impact of neutrino free-streaming on structure formation is thus included in our power spectra. However, since we consider feedback variations at fixed $\sum m_\nu$, the omission of neutrinos from the tracer fields used in constructing our power spectra does not significantly affect our results.}
The total matter fluctuations (indexed with $m$) can be written as a sum of contributions from cold dark matter ($c$) and baryons ($b$):
\begin{equation}
\delta_m = f_c \delta_c + f_b \delta_b.
\label{eq:two_fluid}
\end{equation}

\begin{figure*}
    \centering
    \includegraphics[width = \textwidth]{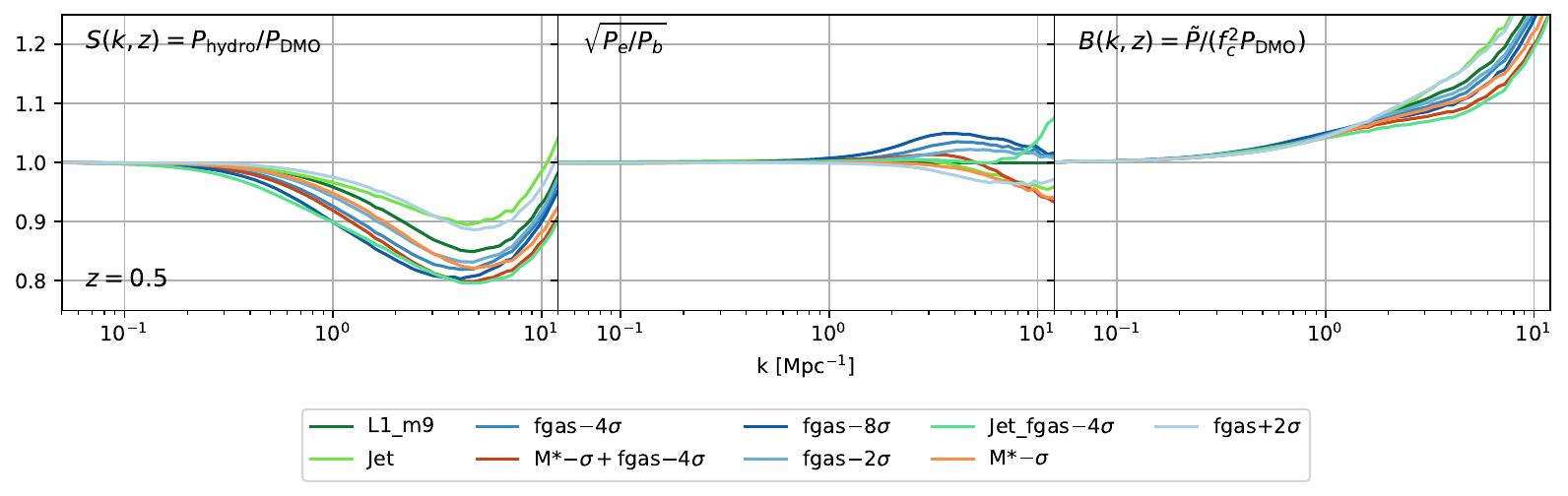}
    \caption{The sensitivity of power spectra to baryonic feedback at $z = 0.5$. Left: $S(k,z=0.5)$ for various feedback models. 
    Centre: the electron to baryon bias. 
    Right: Same as the left, but for the baryon-nulled power spectrum using electrons as tracers of the baryons, as defined in Equation~\ref{eq:ptilde_approx}. 
    While the various feedback models make predictions for the matter power spectrum that vary by $\sim$10\% at $k=1$\,Mpc$^{-1}$, baryon nulling reduces the spread between the curves by almost an order of magnitude at $k \sim 1$ Mpc$^{-1}$.}
    \label{fig:bias}
\end{figure*}

Starting from Eq.~\ref{eq:two_fluid}, the matter power spectrum and matter-baryon cross-power can be written in terms of dark matter and electron power spectra if the electrons (indexed with e) are unbiased tracers of baryons, that is,
\begin{equation}
\delta_e \approx \delta_b 
\label{eq:electrons_trace_baryons}
\end{equation} 
Later, we show this to be the case (see central panel of Fig.~\ref{fig:bias}): the cosmic baryon budget~\citep{fukugita2004cosmic} implies that $\lesssim 2\%$ of the total matter inventory consists of stars and neutral gas.

The lensing autocorrelation is sensitive to $P_{m}$, while lensing x FRB cross correlations are sensitive to $P_{mb}$. In terms of Eq.~\ref{eq:two_fluid} and~\ref{eq:electrons_trace_baryons}, these can be written

\begin{align}
    P_{mm} = f_c^2 P_{c} + &2 f_c f_b P_{cb} + f_b^2 P_{b} \label{eq:pmm}\\
    \approx f_c^2 P_{c} + &2 f_c f_b P_{ce} + f_b^2 P_{e} \label{eq:pmm_approx}; \\
    P_{mb} = f_c P_{cb} + &f_b P_{b} \label{eq:pmb}\\
    \approx  f_c P_{ce} + &f_b P_{e}.\label{eq:pmb_approx}
\end{align}
To null the cross-correlation term in Eq.~\ref{eq:pmm_approx}, we define $\tilde{P}_c$ as
\begin{align}
    \tilde{P}_c &\equiv P_{m} - 2 f_b P_{mb} \label{eq:ptilde}\\
    &= f_c^2 P_{c} - f_b^2 P_{b}.
\end{align} 
$\tilde{P}_c$ is close to $f_c^2 P_{c}$, the true dark matter power spectrum, because $\Omega_b^2 / \Omega_c^2 \approx 4\%$. We can estimate $\tilde{P}_c$ using electrons in place of baryons. We denote this estimate as $\tilde{P}$.
\begin{align}
    \tilde{P} \equiv P_{m} - 2 f_b P_{me}\label{eq:ptilde_approx}
\end{align}

Here, we explore the sensitivity of $\tilde{P}$ to feedback using power spectra extracted from FLAMINGO, a suite of hydrodynamical simulations carried out in a (1 Gpc)$^3$ box. We refer the reader to ~\citet{schaye2023flamingo} and~\citet{kugel2023flamingo} for a complete description of the simulation suite, its eight different feedback variants, and calibration of the subgrid models used therein.
Collectively, the feedback variations in FLAMINGO span a wide range of baryonic feedback models as allowed by the observed galaxy stellar mass function and galaxy cluster gas fractions. 
Notably, the calibration procedure is fully independent of FRBs.

FLAMINGO provides an ideal testbed for evaluating baryon-nulling methods against feedback scenarios.
For instance, the fiducial variant of FLAMINGO is calibrated to be consistent with observational constraints on stellar and AGN feedback: the former via the observed galaxy stellar mass function; and the latter via gas fractions inferred from X-ray observations of groups and clusters~\citep{kugel2023flamingo}. 
The various feedback models in FLAMINGO span a factor of $\sim2$ range ($0.05 < f_{\rm{gas}} < 0.10$) in median gas fraction for haloes at $M_{\rm{500c}} \approx 10^{14} \, \rm{M}_\odot$. In addition, FLAMINGO includes AGN variations which allow thermal and AGN jet-like feedback to be compared; the latter is thought to interact with stellar feedback and expels gas to potentially larger radii~\citep{cielo2018agn, anglesalcazar2017black}.

We extract power spectra out of the FLAMINGO snapshots using the folding technique~\citep{jenkins1998evolution,borrow2020swiftsimio,2021arXiv210605281B} to process wavenumbers $k = 0.02-100 h$ Mpc$^{-1}$. 
Using snapshots at redshifts spanning $z = 0-2$, we measure the power spectra of total matter, dark matter, electrons, and baryons, 
which are defined in the simulations as gas, stars, and black holes (see earlier footnote about neutrinos). We also measure cross-powers between each pair of fields. Then, the power spectra are interpolated as a function of $z$.

The power spectra at $z = 0.5$ referenced to a dark-matter-only counterpart
, are shown in the left panel of Fig.~\ref{fig:bias}. There, we plot $S(k,z) \equiv P_\mathrm{hydro}(k,z) / P_\mathrm{DMO}(k,z)$, showing that the matter power spectrum is suppressed by 10-20\% beyond $k \sim 1$ Mpc$^{-1}$ within FLAMINGO, with $\sim 10\%$ variations across feedback models, each indicated with a different color. $S(k,z)$ within the FLAMINGO suite of simulations is more fully discussed in~\citet{schaller2025flamingo}.

The center panel of Fig.~\ref{fig:bias} shows the electron to baryon bias,
defined as $\sqrt{P_{e}(k) / P_{b}(k)}$. The electron distribution is an
unbiased tracer of baryons at the few-percent level for $k \lesssim 10$ Mpc$^{-1}$. 
This quantity is sensitive to the standard photo-ionization and cooling prescription~\citep{ploeckinger2020radiative} employed in all of the FLAMINGO feedback variations.
However, it justifies the bedrock assumption -- made in this work and in others using the kinetic Sunyaev-Zeldovich (or kSZ) effect -- that electrons trace total baryons on physical scales relevant for weak lensing. 
Recent simulations with more accurate ISM prescriptions~\citep{trayford2025modelling,ploeckinger2025hybrid}, such as the COLIBRE suite~\citep{schaye2025colibre,chaikin2025colibre} will further test this fundamental assumption.

In the right panel, we plot the ratio $B(k,z) \equiv \tilde{P}(k) / (f_c^2 P_\mathrm{DMO}(k))$ at $z = 0.5$.
If dark matter in the hydro-sim was not affected by baryonic feedback, we would expect $B$ curves for all feedback models to be tightly clustered around $B \approx 1$ for a fixed cosmology.

We do see that the spread between $B$ curves is substantially smaller than that of $S(k)$ across all feedback models, and that they are clustered around $B > 1$.
The natural explanation for this is that $B(k,z)$ is not as sensitive to gas physics as $S(k,z)$ because we have nulled the matter-electron cross power $P_{ce}$: of the three terms in Eq.~\ref{eq:pmm_approx}, while the electron auto-power $P_{ee}$ is thought to be the most sensitive to feedback effects~\citep{nicola2022breaking}, the cross-power term consists of the largest mass fraction of the baryonic feedback signal. 

The fact that $B > 1$ indicates that dark matter clustering in the hydrodynamical runs is enhanced with respect to dark matter only runs.
This effect has been dubbed dark matter ``back-reaction'' (hence choosing the letter $B$).
The origin of the enhancement is unclear but has been seen in prior simulations~\citep{van_daalen2011effects}; it may reflect the weak coupling of dark matter to gas expulsion processes which activate at low redshift.
Nevertheless, predictions for $B(k,z)$ are less sensitive to gas physics uncertainties than predictions for $S(k,z)$.
This could reduce the stringent tolerances on modeling and marginalizing over feedback in the next generation of hydrodynamical simulations, and could potentially shrink uncertainties on cosmology since it is insensitive to various feedback models.

Now, we show how to probe $\tilde{P}(k)$ using angular power spectra, in analogy to $P(k)$.
Like $P(k)$, $\tilde{P}(k)$ is not directly accessible observationally. $P(k)$ gets filtered through the lensing kernel and is observed via the cosmic shear angular power spectrum $C^\gamma_\ell$. We define $\tilde{C}^{\gamma}_\ell$ to be $\tilde{P}(k)$ filtered through the same kernel. We now show that $\tilde{C}^\gamma_\ell$ is a linear combination of $C^\gamma_\ell$ and the shear-dispersion cross power $C^{\gamma D}_\ell$, measured by shear autocorrelations and cross-correlations between shear catalogs and FRB dispersion measure catalogs respectively. 

The main challenge is that shear and dispersion are affected by slightly different line-of-sight kernels, which we now show how to account for. Let our tracers ($D$ and $\gamma$) be denoted by superscripts and their underlying fields (electrons $e$ and total matter $m$) be designated by subscripts. Then the angular power spectrum between tracers X and Y as a function of multipole $\ell$, for the underlying fields $x$ and $y$, can be written as an integral over co-moving distance ($\chi$):
\begin{align}    
    C^{XY}_\ell &=\int_{\chi_\mathrm{min}}^{\infty} W_X(\chi) W_Y(\chi) P_{xy}(k = \ell / \chi,z) / \chi^2 ~d\chi. \\
    &= \int_0^{k_\mathrm{max}} W_X(\ell/k) W_Y(\ell/k) P_{xy}(k, z)~\dfrac{dk}{\ell}.
\end{align}
where the integral is cut off at some $k_\mathrm{max} = \ell / \chi_\mathrm{min}$ large enough to not affect the result. $W_D$ and $W_\gamma$ are the window functions for the dispersion measure and shear respectively (see Supplemental Material).

To derive a method to account for the different window functions of dispersion and shear, we make the approximation that the power spectra factorize into functions of $k$ and $z$, i.e.
\begin{equation} 
P_{xy}(k, z) \approx P_{xy}(k) G_{xy}(z) \label{eq:rank1}
\end{equation}
for some \msbold{arbitrary and tracer-dependent growth function} $G_{xy}(z(\chi = \ell / k))$. \msbold{Here, $G_{xy}$ is meant to approximate, as a function of $xy$, the redshift evolution of the power spectrum $P_{xy}(k)$, which in the linear regime evolves for all tracers as $D^2_+(z)$~\citep[see Eq. 8.77 for $D_+$ in][]{dodelson2020modern}.} 

While Eq.~\ref{eq:rank1} does not strictly hold on the quasi-linear scales probed by weak lensing, we will show that the resulting procedure still works on a full-hydro simulation with non-linear, scale-dependent growth. Using~\ref{eq:rank1}, $G_{xy}(\ell/k)$ can be absorbed into the window function
\begin{align}
C^{XY}_\ell = \int_0^{k_{max}} &W_X(\ell/k) W_Y(\ell/k) \times \\
&G_{xy}(z(\chi=\ell/k)) P_{xy}(k)~\dfrac{dk}{\ell}.\label{eq:limber_rank1}
\end{align}
and the integral in Eq.~\ref{eq:limber_rank1} may be discretized into $k$-bins $k_j$ for $\ell$-bins $\ell_i$. 
The angular bandpowers, which are now vectors $c^{XY}_i$ and ${c}^{XY}_j$, can be written as a Riemann sum over the discretized power spectra $p^{xy}_j$ with bin widths $\Delta k_{j}$. We choose the bin widths in $k$ logarithmically with $\Delta k_j \propto k_j$ such that all terms in Eq.~\ref{eq:limber_rank1} except $P_{xy}$ are purely a function of $\ell_i / k_j$. The power spectra can be written
\begin{align} 
{c}^{XY}_i &= \sum_j K^{XY}_{ij} {p}^{xy}_{j} \label{eq:riemann}\\
\intertext{where the matrix $K^{XY}_{ij}$ is Toeplitz~\citep{gray2005toeplitz} and given by}
[K^{XY}_{\ell k}]_{ij} &\propto W_X(\ell_i/k_j) W_Y(\ell_i/k_j) G_{XY}(z(\ell_i/k_j)) \dfrac{k_j}{ \ell_i} \label{eq:matrix_elements}.
\end{align}

\msbold{For brevity we drop the double superscript for auto-correlations (e.g. we write $K^{XX}_{\ell k}$ as $K^X_{\ell k}$, $C^{XX}$ as $C^X$, and so forth).} Then, in matrix notation, the angular power spectra are
\begin{equation}
    c_i^{\gamma} = K_{ij}^{\gamma} p_{j}^{m}
    \label{eq:ckk}
\end{equation}
and
\begin{equation}
    c_i^{\gamma D} = K_{ij}^{\gamma D} p_{j}^{me}
    \label{eq:ckd}
\end{equation}

Combining Eq.~\ref{eq:ckk},~\ref{eq:ckd}, and Eq.~\ref{eq:ptilde_approx}, we can write $\tilde{C}^\gamma_i$ in terms of observable power spectra by inverting the Limber kernel as done in~\cite{gaztanaga1998testing, eisenstein2001correlations,dodelson2000inverting,dodelson2002three}. Discretizing $\tilde{P}$ into $\tilde{p}^c$, we have
\begin{align} 
\tilde{c}^{\gamma}_i = K^{\gamma}_{ij}&~\tilde{p}_j = c^{\gamma}_i - 2 (\Omega_b/\Omega_m) M_{i i'} c^{\gamma D}_{i'}\label{eq:ctilde}
\intertext{where}
M_{i i'} &= (K^{\gamma})_{ij} ((K^{\gamma D})^{-1})_{ji'}.\label{eq:matrix_definition}
\end{align}

\begin{figure}
    \includegraphics[width=0.5\textwidth]{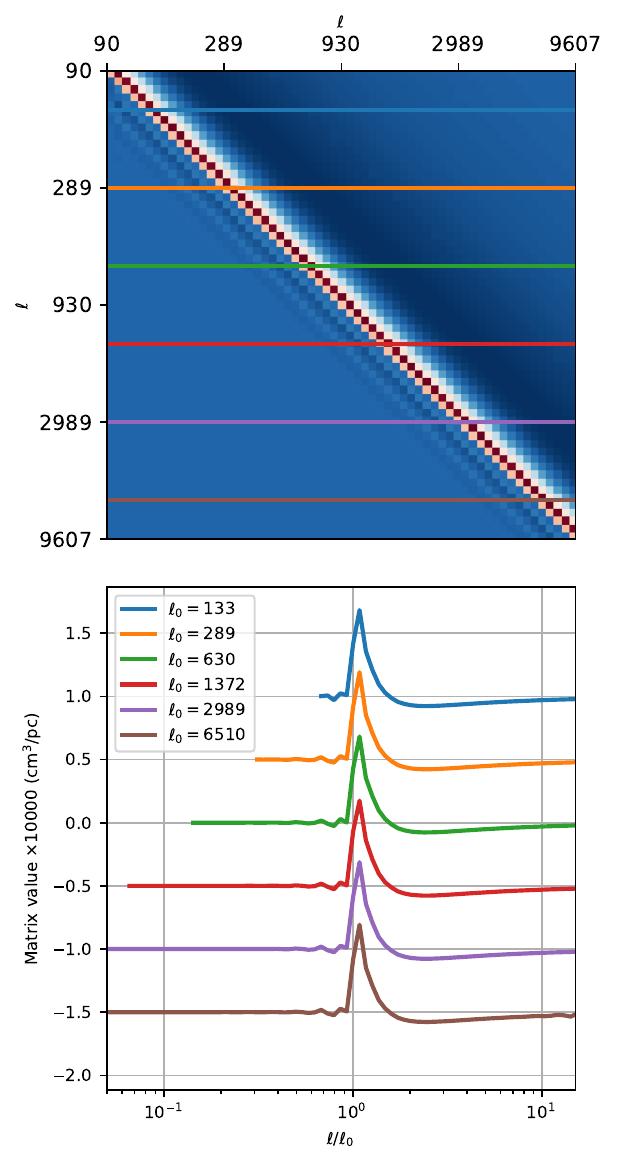}
    \caption{Top: The full re-mixing matrix which converts the FRB dispersion kernel into the lensing kernel evaluated at $\langle z \rangle = 0.5$. Bottom: Individual rows of the matrix $M_{i i'}$ demonstrate that $M_{i i'}$ acts like a smoothing kernel in $\log \ell$. }
    \label{fig:matrix}
\end{figure}
The re-mixing matrix $M_{i i'}$ converts the mode-mixing encoded in the dispersion line-of-sight integral into that of the lensing line-of-sight integral. 
We emphasize that it is agnostic of baryonic feedback and relies only on the Limber approximation and the window functions for weak lensing and FRB dispersion measures, which depend in turn on the mean redshift distribution of galaxy shapes and FRB dispersion measures. 
Numerical experiments confirm that $M_{i i'}$ depends only on the ratio of $G_{me}(z) / G_{m}(z)$, not the exact redshift evolution functions chosen. For simplicity, we choose
\begin{equation} 
    G_{m}(z) / G_{me}(z) = 1,\label{eq:dmm_dme}
\end{equation}
such that the redshift evolution of the $P_{mm}$ and $P_{me}$ are identical. 
Fig.~\ref{fig:matrix} shows the re-mixing matrix $M_{i i'}$ for $\langle z \rangle = 0.5$. 
Using Eq.~\ref{eq:ckk},~\ref{eq:ckd},~\ref{eq:ctilde}, we calculate angular power spectra $c_i^\gamma$, $c_i^{\gamma D}$, and $\tilde{c}_i^{\gamma}$. While realistic measurements must contend with the full shape of $p(z)$~\citep{amon2022dark}, we assume the redshift distributions of FRBs and galaxy shapes are identical top-hat distributions uniformly distributed within $\Delta z = 0.25$ of the mean redshift (here, $\langle z\rangle = 0.5$ and $1$). The power spectra for either case are plotted in Fig.~\ref{fig:exact_lowz} and in the Supplemental Material.
In Fig.~\ref{fig:exact_lowz}, the top row illustrates feedback strongly affects $C^{\gamma D}$, which varies by tens of percent on small scales between feedback variations (note the differing y-axis scales).
Within the fiducial (AGN) feedback mode, $C^{\gamma D}$ correlates positively with higher $f_\mathrm{gas}$, as seen in smaller simulation volumes~\citep{delgado2023predicting}. The correlation also appears in the jet-like AGN feedback variants (``Jet'' versus ``Jet $f_\mathrm{gas} - 4\sigma$'') and in variations with stronger stellar feedback (``$M^*-\sigma$'' versus ``$M^*-\sigma, f_\mathrm{gas}-4\sigma$'').
FRB dispersion measures, like weak lensing shear, are sensitive only to the gas density and distribution rather than its intrinsic properties (e.g. its temperature, pressure, or metallicity). This makes FRBs particularly well-suited for constraining baryonic feedback in cosmic shear studies.

The middle row of Fig.~\ref{fig:exact_lowz} shows the cosmic shear angular power spectra, which are less sensitive to feedback than the FRB cross-correlations.
The final row illustrates the nulled shear $\tilde{C}^\gamma$, which is 2-5 times less sensitive to baryonic feedback than $\tilde{C}$. This holds over a wide range in $\ell$; while we plot a scale cut of $\ell > 1500$~\citep{euclid2020forecast}, a more realistic analysis would include scale cuts tuned as a function of tomographic bin~\citep[e.g.,][]{secco2022dark,robertson2026impact}. 

In the Supplemental Material we show that baryon nulling works at a higher redshift bin ($\langle z \rangle = 1.0$). The reduced spread between different realistic feedback models shows that our approximations 
(Eq.~\ref{eq:electrons_trace_baryons},~\ref{eq:rank1},~\ref{eq:dmm_dme}) introduce systematics smaller than those introduced by realistic feedback models. 

In conclusion, we have introduced baryon nulling: a first-principles method which uses FRBs to null the impact of baryonic feedback on cosmic shear power spectra, and have shown that it performs well at $ z = 0.5$ and $1$ across all feedback models within FLAMINGO.

Our nulling method is model-independent, which makes it complementary to many existing approaches of characterizing baryonic feedback using flexible and accurate forward-modeling. Some methods add baryonic components to dark-matter halo models~\citep{mead2016accurate,mead2020hydrodynamical,tr_oster2022joint,reischke2023calibrating}; others ``baryonify'' the outputs of dark-matter only N-body simulations at the level of haloes, particles, or maps~\citep{schneider2015new,tr_oster2019painting,huang2019modelling,debackere2020impact,osato2023baryon,liu2025fast,anbajagane2024map,sharma2025field}. 
Other methods produce forward models of power spectra via quantities like $S(k,z)$~\citep{aric_o2021bacco,schaller2025flamingo,foreman2020baryonic,chisari2018impact,chisari2019modelling,schneider2019quantifying,salcido2023spk,salcido2025implications,schaller2025flamingo}, connected via hydrodynamical simulations to gas observables like cluster gas fractions~\citep{khrykin2024cosmic, pakmor2023millenniumtng,salcido2025implications,schaller2025flamingo} or FRB dispersion measures~\citep{medlock2024probing,medlock2025constraining,leung2025stellar,sharma2025hydrodynamical}. 
Accurate hydrodynamical modeling is especially crucial for X-ray and thermal Sunyaev-Zeldovich (tSZ) measurements, which require modeling the gas thermodynamics~\citep{shirasaki2020probing,lau2025cosmology,tr_oster2022joint,pandey2022cross,s_anchez2023mapping,pandey2025constraints}. 
To reduce reliance on forward modeling, model-independent approaches like those applied to kSZ+tSZ observations~\citep{amodeo2021atacama,schaan2021atacama,hadzhiyska2024evidence,mccarthy2025flamingo} are especially valuable. Like FRB dispersion measures, kSZ observations directly probe the gas density profile, but may have to contend with line-of-sight peculiar-velocity cancellations~\citep{hadzhiyska2026shear}.

In practice, realistic binning with lensing tomographic techniques~\citep{hu1999power,jain2003cross,huterer2005nulling} like the BNT transform~\citep{bernardeau2014cosmic,taylor2018kcut,fronenberg2023new}, or cross-correlations with galaxies~\citep{sharma2025probing} split into $\approx 10$ redshift bins are likely to strengthen our conclusions. Recent measurements of the FRB-galaxy cross power spectrum~\citep{wang2025measurement} have obtained a detection in several tomographic bins and are particularly encouraging.

Independent forecasts~\citep{reischke2023calibrating} suggest that for a Euclid-like survey, $\approx 10^4$ FRBs are required to constrain baryonic feedback in cross-correlations with cosmic shear. 
A detection is well within reach of the next generation of funded FRB surveys~\citep{vanderlinde2019canadian,hallinan2019dsa}. 
Nevertheless, baryon nulling will be less useful if the method introduces excessive statistical uncertainties. A full forecast is out of the scope of this work, but we estimate the number of FRBs required per galaxy shape such that baryon nulling does not dominate the uncertainties in $\tilde{C}_\ell$. The multipole uncertainties\footnote{also note the factor of 2 from cross-correlation versus auto-correlation statistics.} are

\begin{align}
    (\Delta C^{\gamma}_\ell)^2 \approx&  
        \frac{2}{f_\textrm{sky} (2 \ell +1)}
        \left(
            C^{\gamma}_\ell + \frac{\sigma_\epsilon^2}{\bar{n}_\epsilon}
        \right)^2,\label{eq:poisson_kk}
        \\
        \intertext{and}
            (\Delta C^{D\gamma}_\ell)^2 \approx&
        \frac{1}{f_\textrm{sky} (2 \ell +1)}
        \left(
        C^{D\gamma}_\ell + \frac{\sigma_D^2}{\bar{n}_D}
        \right)^2.
        \label{eq:poisson_kd}
\end{align}
where $\sigma_\epsilon^2 \approx 0.3^2$ and $\sigma_D^2$ (the host DM variance) is $\approx$(100\,pc/cm$^{3})^2$~\citep{madhavacheril2019cosmology}; $\bar{n}_\epsilon$ and $\bar{n}_D$ are the 2D density of shape and DM measurements respectively.

Neglecting the cosmic variance terms, which are subdominant at the small angular scales of interest, the criterion for nulling to not dominate the noise is
\begin{equation}
\Delta C^{D\gamma}_\ell < \Delta C^{\gamma}_\ell / (2 f_b S),
\label{eq:noise_criterion}
\end{equation}
where $S \equiv C^{\gamma}_{\ell = 100} / C^{D\gamma}_{\ell = 100}$ is the
scaling factor that accounts for the different constants and
normalizations of the cosmic shear and dispersion measure power spectra in the linear regime where $P_{mm} \propto P_{mb}$. At $\langle z \rangle = 0.5$, we find $S^{-1} \approx 1.4 \times 10^{4}$ pc/cm$^{3}$ and 7700 pc/cm$^{3}$ at $\langle z \rangle = 1.0$.
Combining Eq.~\ref{eq:poisson_kk} and~\ref{eq:poisson_kd} we have
\begin{equation}
    \frac{\bar{n}_D}{\bar{n}_\epsilon} >
        2 f_b^2 S^2
        \frac{\sigma_D^2}{\sigma_\epsilon^2}.
\end{equation}

Plugging in values for relevant future surveys, we find that nulling will not dominate the noise provided we have one FRB in the overlapping sky region for every $2.4\times 10^{4}$ $~(7\times 10^{3}$) galaxy shapes at $z = 0.5~(1.0)$. The number of galaxy shape measurements expected from Euclid is $30$/arcmin$^{2}$, corresponding to required FRB number densities of 4 (14) FRBs per square degree at $z = 0.5~(1.0)$. While this number is sensitive to the assumed $\sigma_D^2$, which may vary as a function of the properties of the selected FRB host galaxy sample~\citep{bernales2025empirical,leung2025stellar}, our fiducial value of $\sigma_D^2$ is consistent with state-of-the-art population models~\citep{shin2023inferring}. This is easily within reach of funded surveys like CHORD~\citep{vanderlinde2019canadian} and the DSA~\citep{hallinan2019dsa}, which will collectively detect $\mathcal{O}(20,000)$ bursts per year over the Northern sky starting in the next several years. One may also look forward to sensitive expansions of all-sky monitors like BURSTT~\citep{lin2022burstt}, which may boost detection rates by another order of magnitude. All of these FRB surveys cover the footprint of Euclid's Wide Area Survey and overlap with Rubin at the equator. The DSA is particularly well-positioned for weak lensing FRB cross-correlations in the Rubin era because of its significant coverage of southern declinations~\citep{hallinan2019dsa}. 


\begin{figure*}
    \centering
    \includegraphics[width=\linewidth]{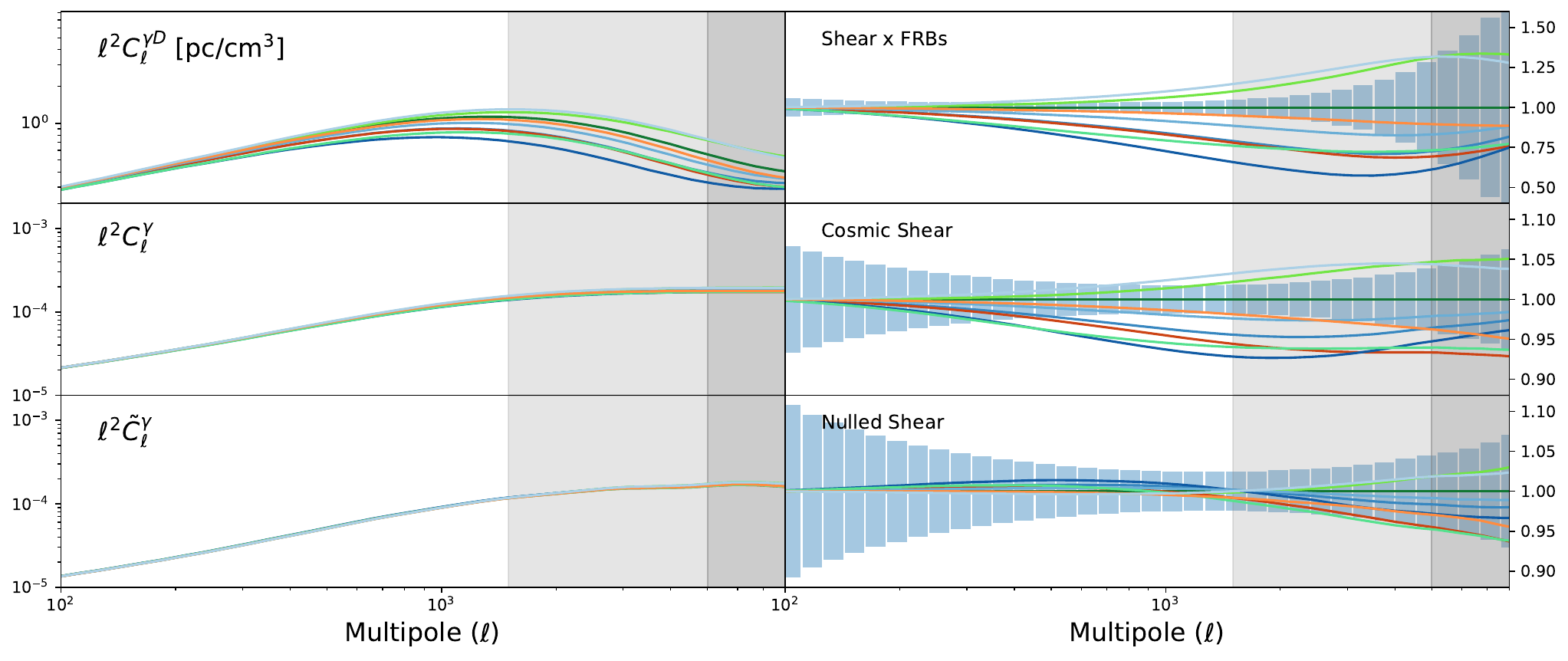}
    \caption{Angular power spectra $C_{\ell}$ as a function of multipole $\ell$ using the $\langle z \rangle = 0.5$ redshift bin. Left column: in the top, middle, and bottom row respectively, we plot the shear-dispersion cross power, cosmic shear auto-power, and the nulled power spectrum ($C^{\gamma D}, C^{\gamma}$, $\tilde{C}^{\gamma}$) for each feedback variation colored identically to Fig.~\ref{fig:bias}. Right column: To show the impact of feedback, we have divided each curve in the left column by the fiducial variant. The light and dark gray bands indicate the conservative and aggressive scale cuts adopted by~\citet{euclid2020forecast}, but depending on the analysis method and tomographic binning, different scale cuts may be appropriate~\citep{taylor2018kcut,taylor2021xcut}. Through baryon nulling, the impact of feedback for all feedback models is reduced by a factor of several within the light gray band and virtually eliminated at larger scales. Uncertainties (Eq.~\ref{eq:poisson_kk},~\ref{eq:poisson_kd}) are shown for $n_D = 10/\textrm{deg}^2$; $n_\epsilon = 30/\textrm{arcmin}^2$; $f_{sky} = 0.5$, though we caution that uncertainties on $\tilde{C}^{\gamma}$ are correlated.}
    \label{fig:exact_lowz}
\end{figure*}
Like in standard weak lensing analyses, a realistic analysis will need to marginalize over redshift distributions of the sources (FRB DMs and galaxy shapes) and feedback parameters. The universal baryon fraction $f_b$ also needs to be marginalized over, but it is already constrained precisely from FRB observations~\citep{macquart2020census,connor2025gas}. In practice, our nulled observables have similar fractional sensitivity to cosmological parameters as existing cosmic shear analyses, but the RMS scatter in $\tilde C_\ell$ over the nine feedback variations is $0.4\%$ at $\ell = 973$ corresponding to $K_{max}=0.5$ as suggested in~\citep{robertson2026impact}, improved by about an order of magnitude from the RMS scatter in $C_\ell$. This precision is likely more than sufficient for precision $S_8$ cosmology and opens up the possibility for weak lensing to access modified dark-sector physics at $k >> 0.1$ Mpc$^{-1}$: theory predictions for $B(k,z)$ may readily be computed and applied to existing hydrodynamical simulations, e.g. those featuring different neutrino masses~\citep{reischke2023calibrating}, axions in large-scale structure~\citep{preston2025prospects}, or primordial non-Gaussianity~\citep{ballardini2024euclid}. 

\begin{acknowledgments}
\emph{Acknowledgments:} C. L. would like to thank Simone Ferraro, Boryana Hadzhiyska, Emmanuel Schaan, Robert Reischke, Steffen Hagstotz, Jaime Salcido, Marcel van Daalen, and various members of the FLAMINGO collaboration for insightful discussions and comments on this work. C. L. is supported by the Miller Institute for Basic Research. 
\end{acknowledgments}

\bibliography{all_combined.bib}

\end{document}